# Fractal space-time fluctuations: A signature of quantumlike chaos in dynamical systems

## A. M. Selvam

Deputy Director (Retired)

Indian Institute of Tropical Meteorology, Pune 411 008, India

Email: amselvam@eth.net

Web sites: http://www.geocities.com/amselvam

http://amselvam.tripod.com/index.html

## Abstract

Dynamical systems in nature such as fluid flows, heart beat patterns, rainfall variability, stock market price fluctuations, etc. exhibit selfsimilar fractal fluctuations characterized by a zigzag (an increase followed by a decrease and vice-versa) curve on all scales in space and time. Power spectral analyses of fractal fluctuations exhibit inverse power law form indicating long-range space-time correlations, identified as self-organized criticality. Mathematical models for simulation and prediction of dynamical systems are nonlinear so that analytical solutions are not available. Finite precision computed solutions are sensitively dependent on initial conditions and give chaotic solutions, identified as deterministic chaos. The author has proposed a general systems theory, which predicts the observed fractal fluctuations as signatures of quantumlike chaos. The model visualizes the irregular fluctuations to result from the superimposition of a continuum of eddies where the dominant component eddies can be resolved using spectral analyses. The model shows that (1) the fractal fluctuations result from an overall logarithmic spiral trajectory with the quasiperiodic Penrose tiling pattern for the internal structure. Conventional power spectral analysis of such a logarithmic spiral trajectory will show a continuum of eddies with progressive increase in phase. (2) Power spectral analyses of fractal fluctuations of dynamical systems exhibit the universal inverse power law form of the statistical normal distribution. Such a result indicates that the additive amplitudes of eddies, when squared (namely the variance) represent the probabilities, a characteristic exhibited by the subatomic dynamics of quantum systems such as the electron or photon. Further, long-range space-time correlations or non-local connections such as that exhibited by macroscale dynamical systems characterize quantum systems also. Therefore selfsimilar fractal fluctuations generic to dynamical systems of all scales in nature is a signature of quantumlike chaos. The model concepts are applied to show that the frequency distribution of bases A, C, G, T in Human Chromosome Y DNA exhibit long-range spatial correlations.

PACS 2003 Classification codes: 89.75.-k: Complex systems; 89.75.Da: Systems obeying scaling laws; 89.75.Fb: Structures and organization in complex systems

*Keywords: chaos, fractals, self-organized criticality, nonlinear dynamics*

## Introduction

Dynamical systems such as rainfall, heart rhythms, aggregate market economics, the electrical activity of the brain, etc., exhibit selfsimilar fractal space-time fluctuations. Such irregular fluctuations may be visualized to result from the superimposition of an ensemble of eddies or sine waves. The power spectra of fractal fluctuations exhibit inverse power law form $f^{\alpha}$ where $f$ is the frequency and $\alpha$ the exponent signifying long-range space-time correlations identified as self-organized criticality (Bak, Tang and Wiesenfeld, 1988). Such non-local connections indicate that for the range of space-time scales for which the slope $\alpha$ of the spectrum is a constant, the amplitudes of the component eddies are related to each other by a scale factor alone independent of the physical, chemical, physiological processes underlying the observed fractal fluctuations of the dynamical system. The exact physical mechanism for the observed self-organized criticality is not yet identified. The author has developed a general systems theory for turbulent fluid flows, which shows that the observed fractal space-time fluctuations of dynamical systems are signatures of quantumlike chaos. The theory provides unique quantification for self-organized criticality in terms of the statistical normal distribution (Selvam, 1990; Selvam and Fadnavis, 1998). The model concepts are applicable to all dynamical systems and in this paper it is shown that the frequency distribution of bases (nucleotides) A, C, G, T in Human Chromosome Y DNA exhibit long-range spatial correlations. The results of the study indicate that the DNA molecule functions as a unified whole self-organized network comprised of the cooperative existence of both noncoding *introns* as well as the coding *exons*. Recent investigations (Kettlewell, 2004; Pearson, 2004) have shown conclusive proof that the noncoding *introns*, also called '*junk DNA*' plays a non-trivial role in the effective functioning of the coding *exons*.

## 2. Model concepts

In summary (Selvam, 1990; Selvam and Fadnavis, 1998), the model is based on Townsend's concept (Townsend, 1956) that large eddy structures form in turbulent flows as envelopes of enclosed turbulent eddies. Such a simple concept that space-time averaging of small-scale structures gives rise to large-scale space-time fluctuations leads to the following important model predictions.

### 2.1 Quantumlike chaos in turbulent fluid flows

Since the large eddy is but the integrated mean of enclosed turbulent eddies, the eddy energy (kinetic) distribution follows statistical normal distribution according to the Central Limit Theorem (Ruhla, 1992). Such a result that the additive amplitudes of eddies, when squared, represent probability distributions is found in the subatomic dynamics of quantum systems such as the electron or photon. Atmospheric flows, or, in general turbulent fluid flows follow quantumlike chaos.

## 2.2 Dynamic memory (information) circulation network

The root mean square (r.m.s.) circulation speeds $W$ and $w_*$ of large and turbulent eddies of respective radii $R$ and $r$ are related as

$$W^2 = \frac{2}{\pi} \frac{r}{R} w_*^2 \tag{1}$$

Eq.(1) is a statement of the law of conservation of energy for eddy growth in fluid flows and implies a two-way ordered energy flow between the larger and smaller scales. Microscopic scale perturbations are carried permanently as internal circulations of progressively larger eddies. Fluid flows therefore act as dynamic memory circulation networks with intrinsic long-term memory of short-term fluctuations.

## 2.3 Quasicrystalline structure

The flow structure consists of an overall logarithmic spiral trajectory with Fibonacci winding number and quasiperiodic Penrose tiling pattern for internal structure (Fig.1). Primary perturbation $OR_O$

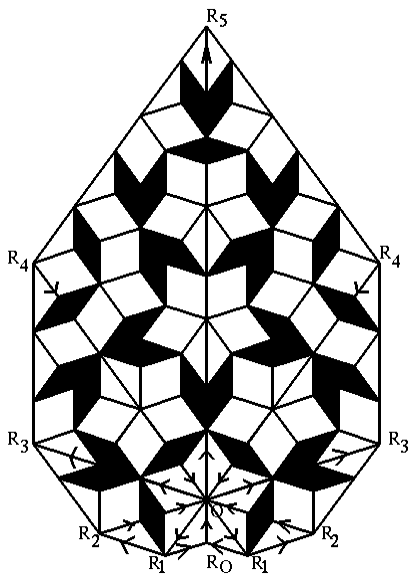

(Fig.1) of time period $T$ generates return circulation $OR_1R_O$ which, in turn, generates successively larger circulations $OR_1R_2$, $OR_2R_3$, $OR_3R_4$, $OR_4R_5$, etc., such that the successive radii form the Fibonacci mathematical number series, i.e., $OR_1/OR_O = OR_2/OR_1 = \ldots\ldots = \tau$ where $\tau$ is the golden mean equal to $(1+\sqrt{5})/2 \approx 1.618$. The flow structure therefore consists of a nested continuum of vortices, i.e., vortices within vortices.

Figure 1: The quasiperiodic Penrose tiling pattern which forms the internal structure at large eddy circulations

## 2.4 Dominant periodicities

Dominant quasi-periodicities $P_n$ corresponding to the internal

circulations (Fig.1) $OR_OR_1$, $OR_1R_2$, $OR_2R_3$, ….. are given as

$$P_n = T(2 + \tau)\tau^n \qquad\qquad (2)$$

The dominant quasi-periodicities are equal to *2.2T, 3.6T, 5.8T, 9.5T,* ……for values of *n = -1, 0, 1, 2,* respectively (Eq.2). Space-time integration of turbulent fluctuations results in robust broadband dominant periodicities (wavelengths), which are functions of the primary perturbation time period *T* alone and are independent of exact details (chemical, electrical, physical etc.) of turbulent fluctuations. Also, such global scale oscillations in the unified network are not affected appreciably by failure of localized microscale circulation networks.

Wavelengths (or periodicities) close to the model predicted values have been reported in weather and climate variability (Selvam and Fadnavis, 1998), prime number distribution (Selvam, 2001a), Riemann zeta zeros (non-trivial) distribution (Selvam, 2001b), Drosophila DNA base sequence (Selvam, 2002), stock market economics (Selvam, 2003), Human chromosome 1 DNA base sequence (Selvam, 2004).

Similar unified communication networks may be involved in biological and physiological systems such as the brain and heart, which continue to perform overall functions satisfactorily in spite of localized physical damage. Structurally stable network configurations increase insensitivity to parameter changes, noise and minor mutations (Kitano, 2002).

## 2.5 Universal spectrum of fluctuations

Conventional power spectral analysis will resolve such a logarithmic spiral flow trajectory as a continuum of eddies (broadband spectrum) with a progressive increase in phase angle.

The power spectrum, plotted on log-log scale as variance versus frequency (period) will represent the probability density corresponding to normalized standard deviation *t* equal to $(\log L/\log T_{50})$ −1 where *L* is the period in years and $T_{50}$ is the period up to which the cumulative percentage contribution to total variance is equal to *50*. The above expression for normalized standard deviation *t* follows from model prediction of logarithmic spiral flow structure and model concept of successive growth structures by space-time averaging. Fluctuations of all scales therefore self-organize to form the universal inverse power law form of the statistical normal distribution.

# 3. Applications of the general systems theory concepts to genomic DNA base sequence structure

DNA sequences, the blueprint of all essential genetic information, are polymers consisting of two complementary strands of four types of bases: adenine (A), cytosine (C), guanine (G) and thymine (T). Among the four bases, the presence of A on one strand is always paired with T on the opposite strand, forming a "base pair" with 2 hydrogen bonds. Similarly, G and C are complementary to one another, while forming a base pair with 3 hydrogen bonds. In addition, the frequency of A(G) on a single strand is approximately equal to the frequency of T(C) on the same strand, a phenomenon that has been termed "strand symmetry" or "Chargaff's second parity". The full story of how DNA really functions is not merely what is written on the sequence of base-pairs; The DNA functions involve information transmission over many length scales ranging from a few to several hundred nanometers (Ball, 2003).

One of the major goals in DNA sequence analysis is to gain an understanding of the overall organization of the genome, in particular, to analyze the properties of the DNA string itself. Long-range correlations in DNA base sequence structure, which give rise to $1/f$ spectra have been identified (Azad *et al.*, 2002). Such long-range correlations in space-time fluctuations is very common in nature and Li (2004) has given an extensive and informative bibliography of the observed $1/f$ noise or $1/f$ spectra, where $f$ is the frequency, in biological, physical, chemical and other dynamical systems.

The long-range correlations in nucleotide sequence could in principle be explained by the coexistence of many different length scales. The advantage of spectral analysis is to reveal patterns hidden in a direct correlation function. The shape of the $1/f$ spectra differs greatly among sequences. Different DNA sequences do not exhibit the same power spectrum. The final goal is to eventually learn the 'genome organization principles' (Li, 1997). The coding sequences of most vertebrate genes are split into segments (*exons*) which are separated by noncoding intervening sequences (*introns*). A very small minority of human genes lack noncoding *introns* and are very small genes (Strachan and Read, 1996).

In the following it is shown that the frequency distribution of Human chromosome Y DNA bases A, C, G, T concentration per 10bp (non-overlapping) follows the model prediction (Section 2) of self-organized criticality or quantumlike chaos implying long-range spatial correlations in the distribution of the bases along the DNA base sequence.

# 4. Data and Analysis

## 4.1 Data

The Human chromosome Y DNA base sequence was obtained from the entrez Databases, Homo sapiens Genome (build 34 Version 2) at http://www.ncbi.nlm.nih.gov/entrez. The ten contiguous data sets containing a minimum of 70 000 base pairs chosen for the study are given in Table 1.

### Table 1: Data sets used for analyses

| Set no | Accession number | Base pairs used for analysis | | Number of data sets |
|--------|------------------|------|------|----------------------|
|        |                  | from | to   |                      |
| 1      | NT_079581.1      | 1    | 70000   | 1   |
| 2      | NT_079582.1      | 1    | 700000  | 10  |
| 3      | NT_079583.1      | 1    | 560000  | 8   |
| 4      | NT_079584.1      | 1    | 350000  | 5   |
| 5      | NT_011896.8      | 1    | 6300000 | 90  |
| 6      | NT_011878.8      | 1    | 1050000 | 15  |
| 7      | NT_011875.10     | 1    | 9870000 | 141 |
| 8      | NT_011903.10     | 1    | 4900000 | 70  |
| 9      | NT_025975.2      | 1    | 70000   | 1   |
| 10     | NT_079585.1      | 1    | 280000  | 4   |

## 4.2 Power spectral analyses: variance and phase spectra

The number of times bases A, C, G, T occur in successive blocks of 10 bases were determined in successive length sections of 70000 base pairs giving frequency distribution series of 7000 values for each data set. A total number of 345 data sets each of length 7000 values for each of the four bases A, C, G, T were available for the study. The power spectra of frequency distribution of the four bases in the data sets were computed accurately by an elementary, but very powerful method of analysis developed by Jenkinson (1977) which provides a quasi-continuous form of the classical periodogram allowing systematic allocation of the total variance and degrees of freedom of the data series to logarithmically spaced elements of the frequency range (0.5, 0). The cumulative percentage contribution to total variance was computed starting from the high frequency side of the spectrum. The power spectra were plotted as cumulative percentage contribution to total variance versus the normalized standard deviation $t$. The statistical chi-square test (Spiegel, 1961) was applied to determine the 'goodness of fit' of variance spectra with statistical normal distribution. Results of power spectral analyses (averages) are given in Table 2. The average variance spectra for the ten contiguous data sets (Table 1) are given in Figure 2.

Table 2. Results of power spectral analyses: averages for the 345 data sets

| base | variance spectra same as normal distribution (%) | average $T_{50}$ in units of 10bp | average frequency/10bp |
|------|--------------------------------------------------|-----------------------------------|------------------------|
| A | 91.6 | 5.74 | 3.00 |
| C | 98.0 | 4.98 | 1.98 |
| G | 97.7 | 5.03 | 1.99 |
| T | 91.6 | 5.75 | 3.04 |

Figure 2: Average power spectra. Continuous lines represent the mean variance spectra and the open circles the corresponding statistical normal distribution

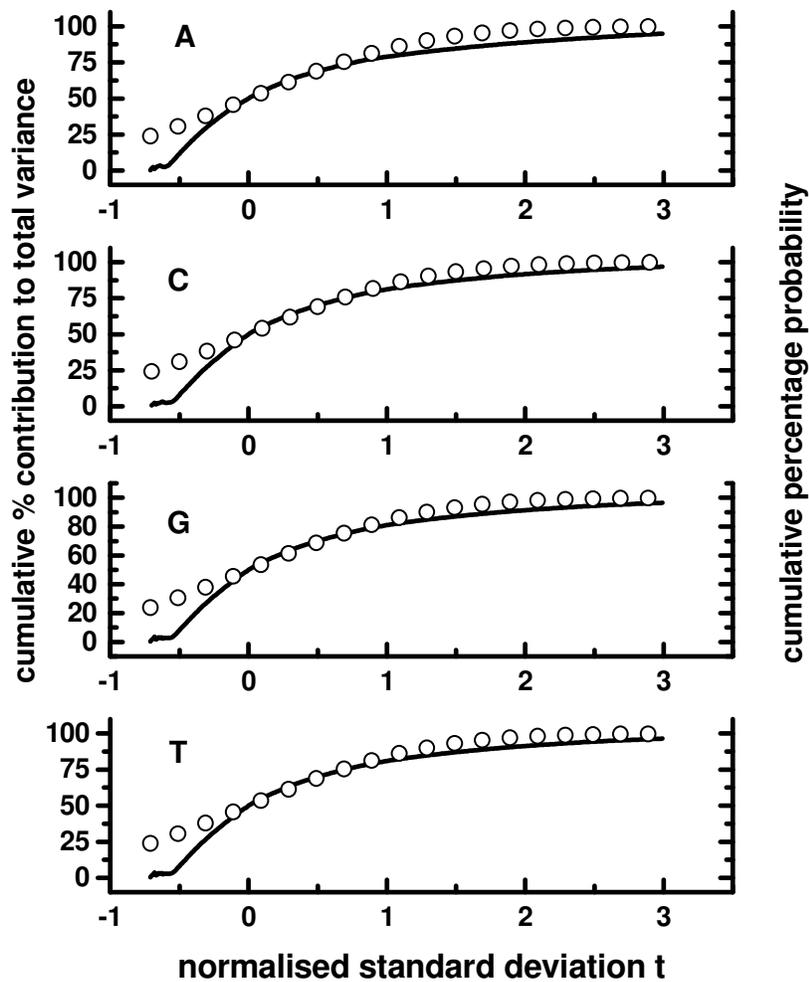

**mean variance spectra for frequency distribution of bases A, C, G, T of human chromosome Y DNA**

# 5. Conclusions

The important conclusions of this study are as follows: (1) the frequency distribution of bases A, C, G,T per 10bp in chromosome Y DNA exhibit selfsimilar fractal fluctuations which follow the universal inverse power law form of the statistical normal distribution (Fig.2), a signature of quantumlike chaos. (2) Quantumlike chaos indicates long-range spatial correlations or 'memory' inherent to the self-organized fuzzy logic network of the quasiperiodic Penrose tiling pattern (Fig.1). (3) Such non-local connections indicate that coding *exons* together with non-coding *introns* contribute to the effective functioning of the DNA molecule as a unified whole. Recent studies indicate that mutations in *introns* introduce adverse genetic defects (Cohen, 2002). (4) The space filling quasiperiodic Penrose tiling pattern provides maximum packing efficiency for the DNA molecule inside the chromosome.

The general systems theory described in this paper is applicable to all dynamical systems including sociology and behavioural sciences. Cor van Dijkum (2001) has stated that mathematical theory of complexity can be used in the social sciences to understand complex social phenomena.

# Acknowledgement

The author is grateful to Dr. A. S. R. Murty for encouragement.